# Synthesis of Heavy Fermion CeCoIn$_5$ Thin Film via Pulsed Laser Deposition

**Jihyun Kim, Hanul Lee, Sangyun Lee, Sungmin Park,* and Tuson Park***

*Center for Quantum Materials and Superconductivity, Department of Physics, Sungkyunkwan University, Suwon 16419, Republic of Korea*

**YoonHee Cho and Hyoyoung Lee**

*Department of Chemistry, Sungkyunkwan University, Suwon 16419, Republic of Korea*

**Won Nam Kang**

*Superconductivity and Thin Film Laboratory, Department of Physics, Sungkyunkwan University, Suwon 16419, Republic of Korea*

**Woo Seok Choi**

*Department of Physics, Sungkyunkwan University, Suwon 16419, Republic of Korea*

CeCoIn$_5$ (Co115) thin films have been grown on Al$_2$O$_3$ (000l) substrates through the pulsed laser deposition (PLD). The films are grown mainly along the *c*-axis, with CeIn$_3$ and In-related alloys. The rock-salt type grains are nucleated, where Co115 grains mixed with excess indium are evenly distributed over the substrate. The electrical resistivity of the films shows a Kondo coherence peak near 47 K and the zero-resistance superconducting state at 1.8 K, which is the first observation in the PLD grown thin films of Co115. The Rietveld refinement of the thin films shows that the *c*/*a* ratio (tetragonality) is suppressed to 1.6312 from 1.6374 of single crystals, which is consistent with the linear relationship between the superconducting transition temperature and tetragonality. The good agreement indicates that the PLD could provide an alternative route to tune the 2D character of the critical spin fluctuations to understand the superconducting pairing mechanism of Co115.

*Corresponding author. E-mail: imsuper007@gmail.com, tp8701@skku.edu

# I. INTRODUCTION

CeCoIn$_5$ (Co115) is one of the representative Ce-based heavy fermion (HF) intermetallic compounds with two closely related compounds of CeRhIn$_5$ [1] and CeIrIn$_5$ [2]. It has attracted significant interest because it provides a good opportunity to systematically explore the interplay between magnetism and unconventional superconductivity. Among various physical properties of CeCoIn$_5$, it has the highest superconducting (SC) transition temperature ($T_c$) of 2.3 K at ambient pressure among the class of the Ce-based HF systems [3]. Also a non-Fermi liquid behavior is evidenced in the temperature dependence of resistivity ($\rho(T)$) from $T_c$ to roughly 10 K, where $n = 1$ in $\rho(T) = \rho_0 + AT^n$. The non-Fermi liquid behavior in Co115 is regarded as arising from the abundant spin fluctuations associated with the magnetic quantum critical point [4, 5].

The study of epitaxial thin films of Co115 is of importance to investigate the effects of low dimensionality and to probe the SC order parameter through Josephson junctions and novel quantum phases emerging at low temperatures and high fields inside the superconducting state [6–8]. Unlike single crystals, however, researches on thin films have been rarely reported owing to the difficulty in their synthesis. Izaki *et al.* first reported 400 nm *c*-axis oriented Co115 thin films deposited on Cr buffered MgO substrate by using molecular beam epitaxy (MBE) [9]. The electrical resistivity of the films revealed the Kondo coherent temperature ($T_{\text{coh}}$) at 30 K and the onset of $T_c$ ($T_{c,\text{on}}$) at 2.2 K with a transition width ($\Delta T_c$) of ~ 0.2 K. These properties of the MBE grown thin films are similar to those of single crystals. Similar results on the Co115 thin film deposited on Al$_2$O$_3$ (0001) through MBE were also reported by Foyevtsov *et al.*, which showed $T_{\text{coh}}$ of 38 K and $T_c$ of 2.0 K with $\Delta T_c$ of 0.15 K [10]. The pulsed laser deposition (PLD) technique has been adopted to deposit thin films on various substrates. Hanisch *et al.* fabricated thin films of Co115 that show $T_c$ of 1.95 K. However, the PLD grown films were polycrystalline and their electrical resistance did not reach zero.

Here, we report the growth of Co115 thin films via the PLD technique, where the structural analysis through the Power X-ray Diffraction (PXRD) and Field Emission-Scanning Electron Microscopy (FE-SEM) images showed that the films are preferentially *c*-axis oriented. Electrical resistivity measurements revealed $T_{coh}$ at 46 K and the zero-resistance state at 1.8 K with $T_{c,on}$ of 2.1 K. These results suggest that PLD is a promising technique for the growth of epitaxial thin films of Co115.

## II. EXPERIMENT

Polycrystal Co115 targets were prepared by the arc melting method, wherein stoichiometric elemental metals (Ce 99.9%, Co 99.995%, In 99.999% are purchased from Alfa) and excess In were melted together on a water-cooled copper hearth in an Ar atmosphere with Ti as a moisture getter. The PXRD analysis showed that Co115 polycrystal targets do not contain any secondary phases, except for a small amount of free indium. To avoid moisture and air influences, high vacuum of $3.5 \times 10^{-7}$ Torr was reached in the chamber before deposition. A LightMachinery IPEX-864 excimer laser with $\lambda = 248$ nm was used to ablate the targets, wherein the laser spot on the target surface was kept within 2 mm × 2 mm, resulting in a laser fluence of ~3 J/cm$^2$. The surface of the *c*-cut Al$_2$O$_3$ (0001) substrate was stabilized by heating it at 750 °C for 60 min using a halogen lamp heater before deposition. The deposition temperature was 600 °C, and the total number of laser shots were 36,000 with a repetition rate of 10 Hz. The thickness of CeCoIn$_5$ films was approximately 900 nm.

The crystal orientation of the film was verified through X-ray diffraction, and the diffraction pattern was carefully refined by the FullProf program. FE-SEM and energy dispersive X-ray spectroscopy (EDX) were used to probe the morphology, chemical ratio and its phase. Electrical resistivity was measured with a standard four-probe method in the physical property measurement system (PPMS) for 1.8–305 K and $^3$He-crystat (Heliox VL) system for 0.25–10 K under magnetic fields up to 9 T.

## III.  RESULTS AND DISCUSSION

Figure 1 shows the XRD data of the Co115 thin film and the refined result through the FullProf program, which shows that the film is mostly in the Co115 phase with $CeIn_3$ and free In impurities. Even though it is difficult to find a distinct lattice match relation between the Co(115) and $Al_2O_3$ (0001) surfaces, the Co115 phase is predominantly (00L) oriented and shows (00L) harmonics such as (001), (002), (003), (004), (005), and (006). In particular, the (003) peak has the full width at half maximum (FWHM) of 0.115, which is smaller than that of other PLD grown thin films [9], thereby indicating that the *c*-axis oriented layer is of high quality. Additional peaks such as (100), (200), (111), (112), (113), (114), (203), (311), and (115) are also observed, but their intensities are much lower than those of the *c*-axis oriented layer.

According to EDX results of the Ce115 thin films, the chemical composition of $CeCoIn_5$ is Ce:Co:In = 1.61:1.00:4.51 for the film grown in the rock-salt type, where the deviations from the expected stoichiometry may be ascribed to many facets of the grains and the detecting area that includes grain boundaries. On the other hand, the Ce115 film grown in the micro disk type is Ce:Co:In = 1.10:1.00:4.98. When combined with the XRD results of the film, the stoichiometry of the micro disk type of film indicates that the film in the rock-salt type has been also grown in a CeCoIn$_5$ phase with proper stoichiometry. FE-SEM images in Fig. 2(a) and (b) show that CeIn$_3$ layers are formed underneath Co115 layers, i.e., CeIn$_3$ layers are directly deposited on the Al$_2$O$_3$ (0001) surface. Co115 microdisks without grain boundary were grown on CeIn$_3$ layers, which is also verified from the EDX results on the regions between the microdisks, as depicted by the white-dot box in Fig. 2(a) and (b). Some Co115 grains depicted by the white dot circle in Fig. 2(c) and (d) are covered with a thin Indium layer, which can be attributed to the excess In in the Co115 target. As shown by white arrows in Fig. 2(c), the submicron-sized indium droplets are irregularly located on top of the Co115 layer. The part in contact with the Co115 layers is partly melted, and the area nearby the contact shows a similar visibility as that of the thin indium blanket on the Co115 layer. These observations strongly suggest that the thin Indium blanket originates from the melting of the indium droplet. The melted indium droplet is locally spread over the Co115 layer, forming a very thin layer, which is ascribed to the surface tension caused by both the indium droplet and the Co115 layer. The fact that the indium layer can be easily brushed out with a cotton swab implies that the layer has a different nature from the bulk with regard to the adhesion between indium and Co115 layers.

The growth of Co115 thin films depends on deposition conditions such as the laser fluence (mJ/cm$^2$), deposition temperature, distance between the target and the substrate, and ambient pressure in the PLD. Co115 films does not grow on the Al$_2$O$_3$ (0001) substrate when the laser energy is less than 80 mJ or larger than 260 mJ, instead, CeIn$_3$ layers, non-identical In related alloys, and individual Ce, Co, and In grains are formed. This implies that it is crucial for the formation of Co115 layer that the number of added atoms, Ce, Co & In ions, by ablation and the stoichiometry must be in optimal condition. In order to grow Co115 films, the deposition temperature should be higher than 550 °C and the target distance should be shorter than 60 mm when 120 mJ is used. Because the deposition window required for obtaining Co115 layers is too narrow, the films are typically grown in islands or mostly become CeIn$_3$ layers with individual Ce, Co, and In grains. This is consistent with the fact that CeIn$_3$ is preferentially formed at 450~750 °C because the eutectic point between Ce and Co is too difficult to be reached [3]. As a result, the energetic Ce, Co, and In ions (~ 30% excess) laid on the Al$_2$O$_3$ (0001) substrate can move to form the Co115 phase with a relatively higher laser energy than that of oxide epitaxial films (~ tens of mJ) even though the ablated atoms cannot move easily as in the In flux because of the low diffusivity caused by the interaction between ablated atoms and the substrate.

The refinement of XRD of the Co115 thin film shows that the Co115 layer has lattice constants of $a$ = 4.6265 Å and $c$ = 7.5470 Å. As shown in Table 1, the lattice constant $a$ is slightly larger than that of the single crystal; whereas, the $c$-axis constant is smaller than that of the single crystal. This implies that the film has undergone slight tensile strain, and the tetragonality of 1.6312 is smaller than that of the single crystal (=1.6374) [11]. The linear relationship between the $T_c$ and tetragonality reported by Sarrao *et al*. [11] suggests that the $T_c$ for our film will become 1.8 K, which is compatible with the measured $T_{c,0}$, as shown in Fig. 3(b). The good agreement between the prediction and experimental results suggests that the 900 nm Co115 film can be considered as a single crystal with the elongated $a$-axis and shrunk $c$-axis. Because the tetragonality can change the distance between Ce ions within the Ce-In layer and the distance between Ce-In layers buffered by Co-In blocks, the PLD technique may provide an alternative route to tune the 2D character of the relevant spin fluctuations, which could guide our understating of the SC pairing mechanism of the unconventional superconductor CeCoIn$_5$.

A conventional four-probe technique was applied to measure the electrical resistivity of Co115 thin films. The temperature dependence of resistivity was measured by PPMS (1.8–305 K) and He$^3$ refrigerator (Heliox VL, 0.25–10 K) with magnetic fields up to 9 T. As shown in Fig. 3(a), the overall temperature dependence of the film, $\rho(T)$, is similar to that of the high-quality single crystal: $\rho(T)$ shows the resistivity minimum at 115.89 K, followed by the resistivity maximum at 46 K (= $T_{\text{coh}}$). Even though $T_{\text{coh}}$ of the thin film is similar to that of single crystals (~ 46 K), the resistivity minimum is much shallower and the resistivity peak is much more rounded than that of single crystals, thereby indicating contributions from other sources. Indeed, a slight kink at 9.8 K is observed from the $CeIn_5$ impurity phase and a sharp reduction in the resistivity is apparent owing to the excess In and/or In-containing alloys. Table 1 compares the SC properties of the single crystal and thin film Co115 (Fig. 3(b)). The zero-resistance temperature of the film is 1.8 K, which is lower than 2.3 K of the single crystal. The SC transition width $\Delta T_c$, a difference between the onset and zero-resistance temperatures of the SC transition, is approximately 0.30 K for the thin film, while ~ 0.21 K for single crystals, thereby indicating that the SC quality of thin films is similar to that of single crystals.

Figure 4(a) shows the temperature dependence $\rho(T)$ of the Co115 thin film under magnetic fields up to 9 T. At 0 T, $\rho$ starts to deviate from the normal state tendency below 3.9 K owing to indium related alloys, undergoes the SC transition below 2.16 K ($T_{c,on}$), and reaches the zero resistance value at 1.8 K ($T_{c,0}$) (Fig. 3(b)). With an increase in the magnetic field, both SC transition temperatures gradually decrease. At 2 T, the indium related feature is completely suppressed, whereas the SC transition owing to the thin film Co115 remains. Figure 4(b) plots the temperature dependences of the upper critical fields ($H_{c2}$) determined by $T_{c,50\%}$ (green squares) and $T_{c,0}$ (purple circles) defined by the midpoint of the normal state and $T_c$ zero, respectively. For comparison, $H_{c2}$ of single crystals for the fields applied parallel and perpendicular to the $c$-axis are plotted as red and blue solid lines, respectively [12]. Dashed lines are the best results from the least-squares fitting of the phenomenological expression $H_{c2}(T) = H_{c2}(0)(1-(T/T_c)^2)^n$, where $n = 0.74$ and 0.73 are obtained for $T_{c,0}$ and $T_{c,50\%}$, respectively. The upper critical field $H_{c2}(0)$ from $T_{c,0}$ ($T_{c,50\%}$) is 5.4 (8.6) T, whose value is between 5.0 T for $H \| c$ and 11.6 T for $H \| a$ of single crystals [12]. The inset to Fig. 4(b) describes the dependence on reduced temperature ($T_c/T_{c,0}$) of the reduced upper critical field ($H_{c2}/H_{c2(0)}$) for $CeCoIn_5$ single crystals (solid lines) and thin films (open symbols). Without regard to the single crystal or thin film, the reduced $H_{c2}$s almost collapses on a single curve as a function of the reduced temperature, thereby indicating that the pair breaking effects by the magnetic field may be similar for both the single and thin film $CeCoIn_5$.

## IV. CONCLUSION

We have successfully synthesized thin films of $CeCoIn_5$ on the $c$-cut $Al_2O_3$ substrate by using the PLD. The analysis of the Rietveld refinement showed that the films are preferentially oriented along the $c$-axis. EDX and SEM studies indicated that $CeCoIn_5$ was deposited through an island growth mechanism with a small amount of $CeIn_3$ and In containing secondary phases. In contrast to previous PLD grown thin films that failed to show zero resistance, the electrical resistivity of $CeCoIn_5$ thin films reported in this work revealed the zero-resistance SC transition at 1.8 K and the resistivity maximum at 49 K, which is popularly known as Kondo coherent behavior. The temperature dependence of the upper critical field is in good agreement with that of high-quality single crystals. Further work to optimize the growth conditions is under progress to grow high-quality epitaxial thin films without secondary phases.

## ACKNOWLEDGMENTS

SP and JK were partially supported by Basic Science Research Program through the National Research Foundation of Korea(NRF) funded by the Ministry of Education (NRF-2016R1A6A3A11935214 and NRF-2018R1D1A1B07051040). SP, JK, SL and TP were supported by National Research Foundation (NRF) of Korea grant funded by the Korean Ministry of Science, ICT, and Planning (No. 2012R1A3A2048816). WSC was supported by NRF-2019R1A2B5B02004546.

Table 1. Comparison of superconducting and structural properties of a single crystal and a thin film of CeCoIn$_5$.

|  | $T_c$ (K) | $H_{c2}$ (0) | $a$ (Å) | $c$ (Å) | $c/a$ |
|---|---|---|---|---|---|
| Single crystal [3, 12] | 2.3 K | $a\parallel$ ~11.5 T [12] <br> $c\parallel$ ~5.25 T [12] | 4.614 | 7.555 | 1.637 |
| Thin film | 1.8 K | 5.4 T | 4.627 | 7.547 | 1.631 |

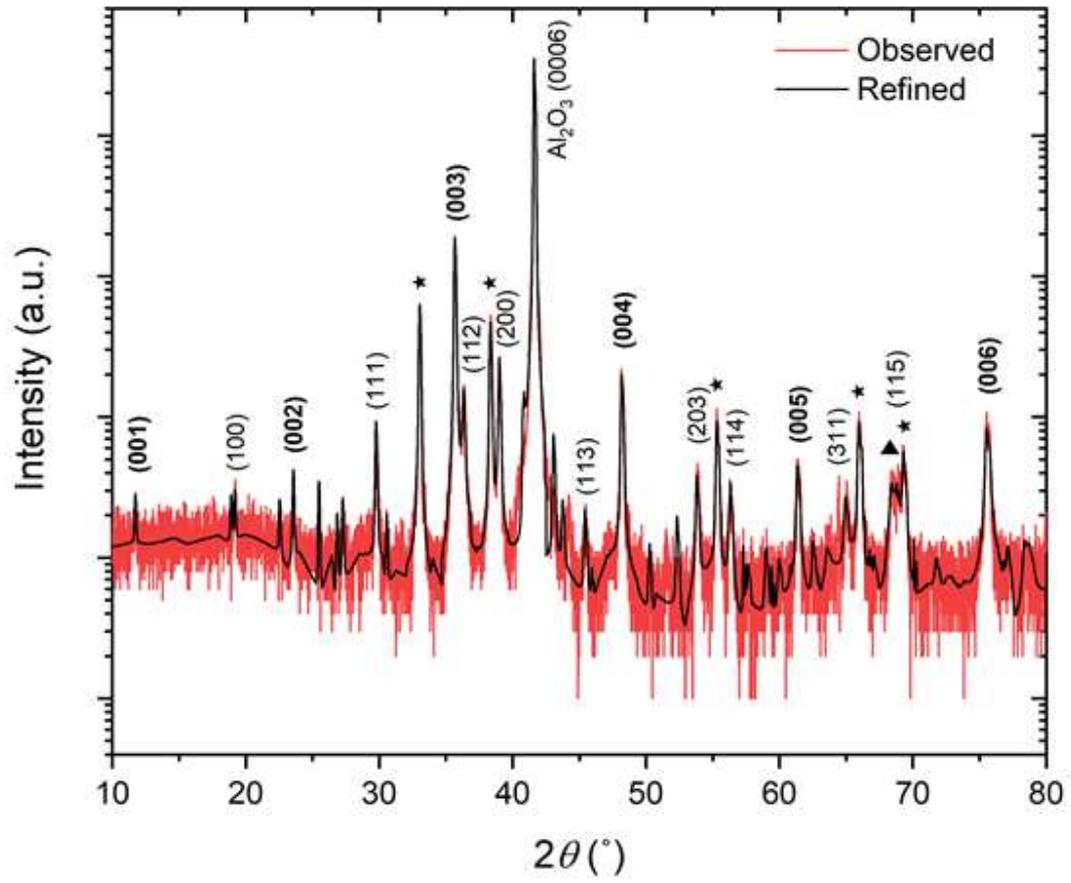

Figure 1 XRD $2\theta$ scan of Co115 thin film on Al$_2$O$_3$ (0001) substrate (*c*-cut). (00L) CeCoIn$_5$ peaks are well observed together with CeIn$_3$(★) and In (▲) peaks. The observed Co115 peaks are described by the red line and are overlapped with the black line refined by Fullprof analysis.

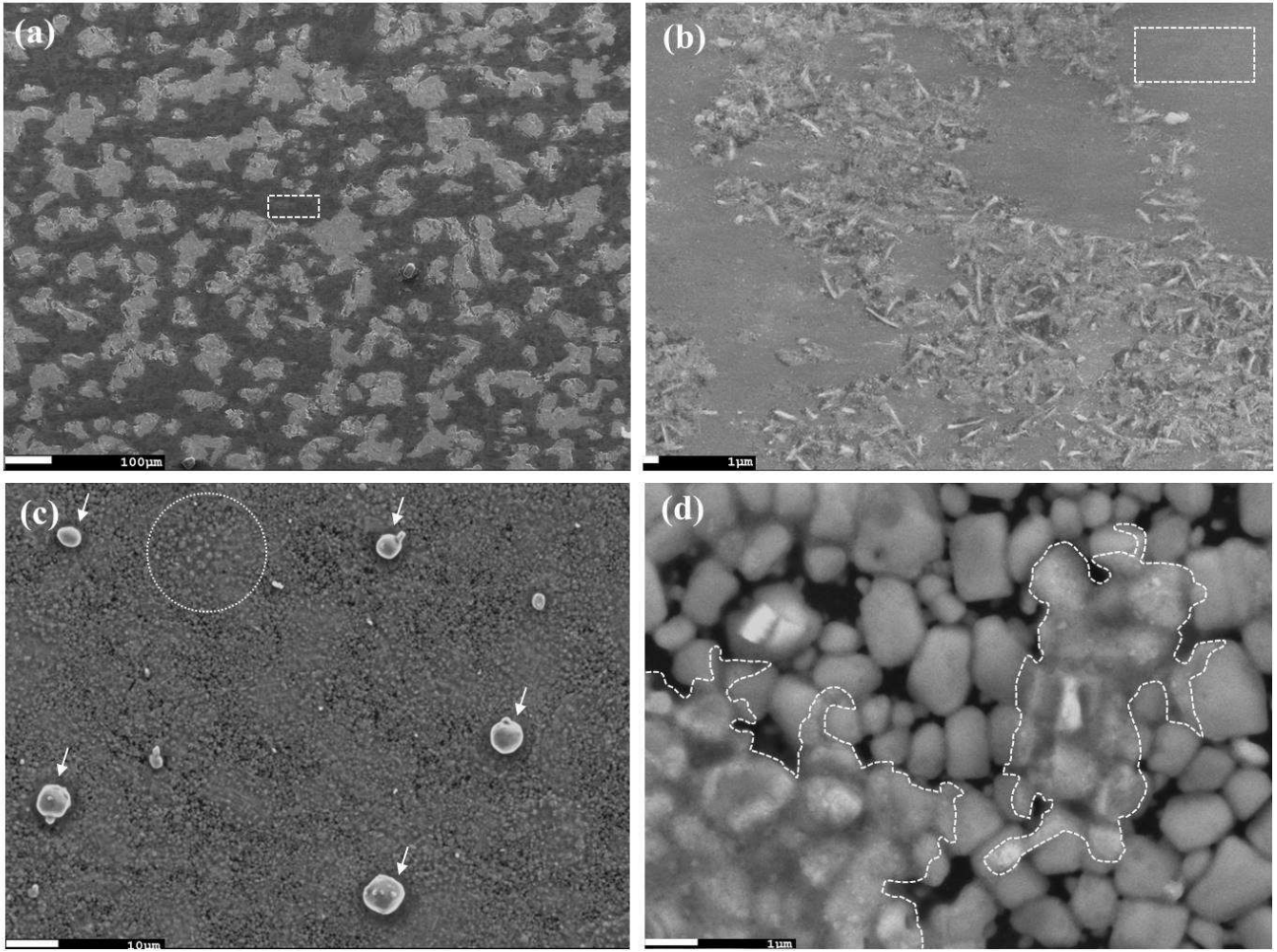

Figure 2 FE-SEM images are shown. (a) CeIn$_3$ layers are observed inside the white dotted box by using EDX analysis. Thin films were grown in microdisks, and CeIn$_3$ layers can be observed in the area at which microdisks were not grown. (b) Some CeIn$_3$ layers are observed in the microdisks of Co115 films, and the scattered needles are Ce alloys with excess indium. (c) The white arrows indicate indium droplets not melted during the deposition process, and the white dot circle describes the region where Co115 grains are covered by indium blankets. (d) The white dot region in (c) is magnified. Very thin indium blankets are indicated by the white dotted boundaries.

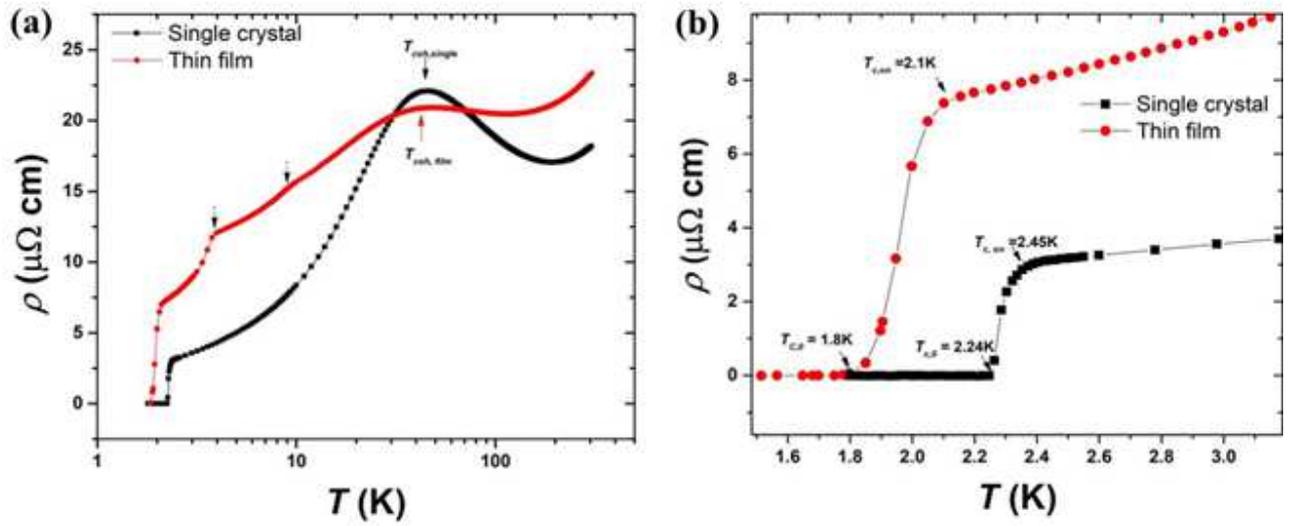

Figure 3. (a) Electrical resistivity is plotted as a function of temperature for single crystal (black symbols) and thin film of CeCoIn5 (red symbols). Resistivity maximum (or Kondo coherence peak) at 46 K (thin film) and 45 K (single crystal) are marked by red and black arrows, respectively. The black-doted arrows at 9.8 and 3.97 K are caused by $CeIn_3$ layers formed underneath Co115 grains. (b) The low $T$ $\rho(T)$ is magnified near the SC transition temperature. The zero-resistance SC transition temperature $T_{c,0}$ is 1.8 and 2.24 K for thin film and singe crystal, respectively.

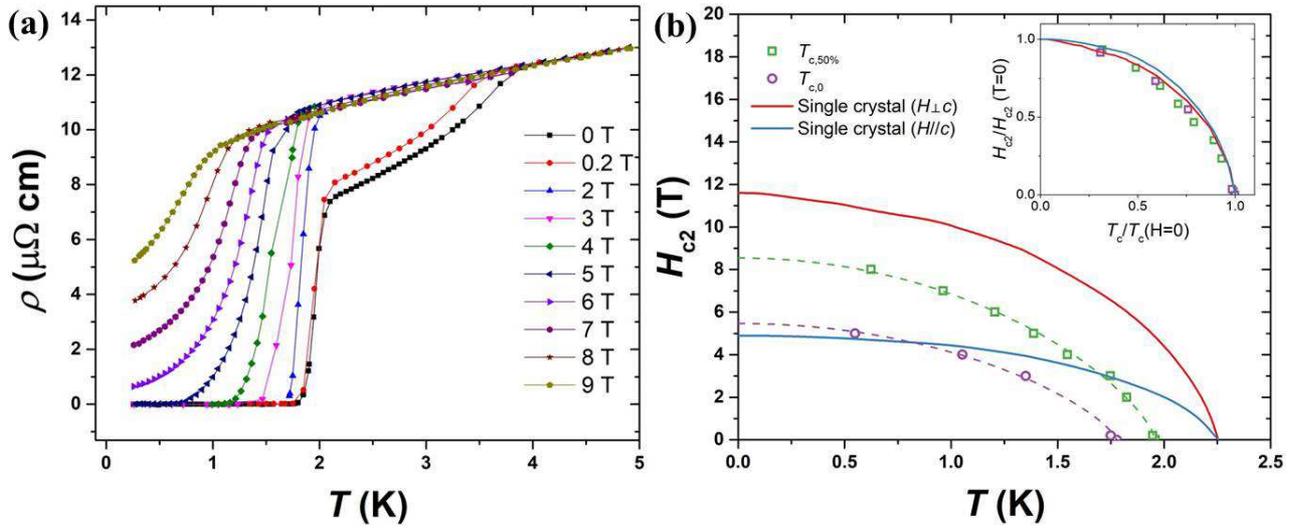

Figure 4. (a) Electrical resistivity of Co115 film is plotted as a function of temperature near SC phase transition under magnetic field up to 9 T. (b) Upper critical field ($H_{c2}$), obtained from the zero-resistance ($T_{c,0}$) and 50% drop ($T_{c,50\%}$) from the normal state, is plotted as a function of magnetic field. The dotted lines are the least-squares fitting of the phenomenological expression $H_{c2}(T)=H_{c2}(0)(1-(T/T_c)^2)^n$. Solid lines are data for $H\parallel c$ (blue line) and $H\parallel a$ (red line) from Ref. [12]. Inset shows the scaling of $H_{c2}/H_{c2}(T=0)$ vs $T_c/T_c(H=0)$.